\documentclass[12pt,a4paper]{article} 
\usepackage{graphicx}
\usepackage{lscape}
\usepackage{amsmath}
\usepackage{subfig}
\usepackage{xcolor}
\usepackage[normalem]{ulem}

\begin{document}

\title{Jovian planets as co-detectors of gravitational waves}
\renewcommand{\thefootnote}{\fnsymbol{footnote}}
\author{\. Ibrahim Semiz\footnote{e-mail: ibrahim.semiz@boun.edu.tr} ~\& A. Kaz\i m \c Caml\i bel\footnote{e-mail: kazim.camlibel@boun.edu.tr}
\\ \small Physics Department, Bo\u gazi\c ci University, \\ \small Bebek, \. Istanbul, Turkey}
\date{ }

\maketitle
\renewcommand{\thefootnote}{\arabic{footnote}}\setcounter{footnote}{0}

\begin{abstract}
Acoustic oscillations in stars can be driven by gravitational waves. However, at present it is not feasible to use the helioseismology data for their detection, since it is impossible to disentangle the uncertain driving contributions originating in the Sun itself.

We here point out that any such wave will affect also Jupiter and Saturn in a similar {\em and coordinated way}; after all, they are mostly spheres of gas like the Sun, only one order of magnitude smaller. Hence, akin to the concept of coincidence detection in particle physics experiments, evaluation of the time-correlation function of the measured acoustic velocities of the same mode of oscillation of any two of these three objects will eliminate the (independent) internal effects; and observation of a peak in that correlation function will be tantamount to detection of a gravitational wave. There is a (slight) possibility that such detection has already occured.
\end{abstract}

The idea that gravitational waves (GW) might affect Earth and other celestial objects has been around since the first attempts, via the Weber bars \cite{weberbook, weberprl} to detect these waves. Dyson \cite{dyson} has written about searching for gravitational wave signatures in seismic data of Earth, and according to him, so did Weber \cite{weber1968}. They decide that the effects are below the noise level.

With the advent of helioseismology \cite{asteroseismologybook, review, review2}, the same considerations for the Sun, and by extension, other stars, became relevant. The response of relativistic stars to gravitational waves seems to be analyzed first by \cite{relativistic}, and nonrelativistic stars by \cite{nonrelativistic1, nonrelativistic2}; see \cite{seismologygw} for a more thorough rewiew of previous work.

This response is important to understand since it will provide the signature to look for in helioseismology data \cite{sbgw1, sbgw}, in other words, it is needed to disentangle the driving contribution of gravitational waves to acoustic solar oscillations from the much larger contributions originating in the Sun itself.

In particle physics experiments, one way to pick up a signal from an overwhelming background is to operate a second similar detector able to detect the {\em same} particles (not just particles of same species), but with independent background. If the detectors have significantly better time resolution then the typical time interval between events, the background detections of the two detectors, not being correlated with each other, will usually not coincide in time, hence can be {\em vetoed}; so-called {\em coincidence detection}.

While we do not live in a binary system, hence do not have a second Sun at our disposal for coincidence detection of gravitational waves, Jupiter and Saturn {\em are} (mostly) spheres of gas like the Sun, only one order of magnitude smaller; hence, it can be expected that they can serve this purpose. In fact, theoretical work \cite{sympa,jupitersaturn,jovianseismology} suggests that Jovian seismology and helioseismology are similar with comparable frequency ranges for global oscillations.

Any gravitational wave impinging on the solar system from the outside will produce similar effects in the Sun, Jupiter and Saturn; up to an hour apart. An impulsive wave might result in a jump each in the amplitude of a certain mode of oscillation for more than one of these objects, the jumps appropriately time-separated; or a long wavetrain could be observed as a peak in the time-correlation function  
\begin{equation}
\int f_{i}(t) f_{j}(t-T) dt \nonumber
\end{equation}
of the acoustic displacements of some pair of these objects.

The feasibility of this enterprise depends on the relative sizes of oscillation amplitudes excited by gravitational waves, oscillation amplitudes excited otherwise, and observational precision. The velocity amplitudes for GW-excited solar oscillations are expected to be $10^{-9}-10^{-3}$ cm/s \cite{seismologygw,sbgw1}, whereas observed solar oscillations have amplitudes up to 10 cm/s \cite{review}, and the SOHO spacecraft \cite{soho} observing the Sun has a sensitivity of 1 mm/s \cite{golf}.  On the other hand, only ground-based observations exist for Jupiter, \textcolor{blue}{by the SYMPA instrument \cite{sympa},} with typical observed amplitude up to 50 cm/s, and sensitivity about 10 cm/s. Of course, longer observation can increase the sensitivity, and possible future orbiters of Jovian planets \cite{juno, juice}  could also collect seismology data \cite{echoes}, further  improving the situation. 

The order of magnitude of the GW contribution to the seismic oscillations may be as much as $10^{-4}$, which might just call for coincidence detection: If the contribution was much larger, i.e. a significant fraction of the observations, it probably would already have been apparent. If on the other hand, if it was much smaller, e.g of the order $10^{-10}$, it would be totally swamped by the turbulent fluctuations in the solar/jovian atmospheres, and be hopeless to detect.

During the writing of this manuscript, we became aware of a comment in \cite{seismologygw} to the effect that observations of oscillations of many stars in the same cluster might help detect a gravitational wave. However, the maximum precision that can be reached by such measurements is about 20 cm/s \cite{asteroseismologybook,astero}, ``with future promise of 1 cm/s''; this precision, and uncertainties in the determinations of stellar distances would seem to make coincidence detection of GWs in this context at the present level of technology very difficult. Furthermore, typical interstellar distances are of the order of light-years, meaning that correlations need to be sought for correspondingly long time intervals. However, the application to binary stars might become feasible in the near future, since they lie much closer to each other than members of a cluster, and the separations of the pairs can be determined accurately without needing similar accuracy for the overall distance.
.

Coincidence detection in the Solar System has another advantage: The Sun and planets can be seen as disks, as opposed to stars, which can only be seen as points. Hence, the observational data can be explicitly projected onto spherical harmonic functions as in Sect. 2.1 of \cite{sympa}, enabling identifications of contributions of many such modes. In asteroseismology, however, one has to use indirect methods, and can only identify a few modes \cite{asteroseismologybook}. 

In fact, there is the off chance that gravitational waves might already have been detected: In \cite{sympa}, unexplained "serendipituous" matches of frequencies of some Solar and Jovian oscillation modes is reported, and even though the SYMPA sensitivity is probably too low to claim this, one cannot help but wonder if excitation by the same gravitational wave might be the explanation.

To summarize, we propose that the Sun and the Jovian planets together can serve as a set of coincidence detectors for gravitational waves impinging on the solar system; negating the need for {\em detailed} modeling of the response of each individual object/detector.

\end{document}